\documentclass[pdflatex,sn-apa]{sn-jnl}% Basic Springer Nature Reference Style/Chemistry Reference Style

\usepackage{graphicx}%
\usepackage{multirow}%
\usepackage{amsmath,amssymb,amsfonts, bm}%
\usepackage{bbm, dsfont}
\usepackage{amsthm}%
\usepackage{mathrsfs}%
\usepackage[title]{appendix}%
\usepackage{xcolor}%
\usepackage{textcomp}%
\usepackage{manyfoot}%
\usepackage{booktabs}%
\usepackage{algorithm}%
\usepackage{algorithmicx}%
\usepackage{algpseudocode}%
\usepackage{listings}%
\usepackage{paralist}
\usepackage{tabularx,longtable}
\usepackage{hyperref}

\setcitestyle{maxcitenames=1}
\newtheoremstyle{thmstyleone}% Numbered
{18pt plus2pt minus1pt}% Space above
{18pt plus2pt minus1pt}% Space below
{\small\itshape}% Body font
{0pt}% Indent amount
{\small\bfseries}% Theorem head font
{}% Punctuation after theorem head
{.5em}% Space after theorem headi
{\thmname{#1}\thmnumber{\@ifnotempty{#1}{ }\@upn{#2}}%
  \thmnote{ {\the\thm@notefont(#3)}}}% Theorem head spec (can be left empty, meaning `normal')
\DeclareMathOperator*{\argmin}{arg\,min}

\begin{document}
%%%%%%%%%%%%%%%%%%%%%%%%%%%%%%%%%%%%%%%%%%%%%%%%%%%%%%%%%%%%%%%%%%
%%%%%%%%%%%%%%%%%%%%%%%%%%%%%%%%%%%%%%%%%%%%%%%%%%%%%%%%%%%%%%%%%%
\title[Article title]{Digital twin-based hybrid framework for steam generator clogging prognostics}

\author[1]{\fnm{Edgar} \sur{Jaber}}
\author*[2,3]{\fnm{Emmanuel} \sur{Remy}\email{emmanuel.remy@edf.fr}}
\author[2,3]{\fnm{Vincent} \sur{Chabridon}}
\author[2]{\fnm{Morgane} \sur{Garo-Sail}}
\author[1]{\fnm{Mathilde} \sur{Mougeot}}
\author[4]{\fnm{Didier} \sur{Lucor}}
\author[5]{\fnm{Jérôme} \sur{Delplace}}
\author[5]{\fnm{Maxime} \sur{Lointier}}

\affil[1]{\orgname{Université Paris-Saclay, CNRS, ENS Paris-Saclay, Centre Borelli}, 
\orgaddress{\postcode{91190}, \city{Gif-sur-Yvette}, \country{France}}}

\affil[2]{\orgname{EDF R\&D}, 
\orgaddress{\street{6 Quai Watier}, \postcode{78401}, \city{Chatou}, \country{France}}}

\affil[3]{\orgname{SINCLAIR AI Lab., Saclay}, 
\orgaddress{\city{Saclay}, \country{France}}}

\affil[4]{\orgname{Université Paris-Saclay, CNRS, Laboratoire Interdisciplinaire des Sciences du Numérique}, 
\orgaddress{\postcode{91405}, \city{Orsay}, \country{France}}}

\affil[5]{\orgname{EDF Nuclear Division}, 
\orgaddress{\street{1 Place Pleyel}, \postcode{93200}, \city{Saint-Denis}, \country{France}}}

%==========================================================%
\abstract{We present a hybrid framework to support prognostics of the clogging degradation phenomenon in tube support plates for digital twins of steam generators in pressurized water reactors. The proposed approach combines a physics-based simulation code, heterogeneous and sparse observational data, and several uncertainty-quantification techniques to obtain a robust estimate of the steam generator remaining useful life associated with the clogging rate. The proposed framework is compatible with a digital twin platform to assist maintenance planning of EDF steam generators.}
%==========================================================%
\keywords{Digital twins, SciML, Clogging, Steam generators, Predictive maintenance, Simulation models, Scarce field data}

\maketitle

\section*{Abbreviations and acronyms}
\label{sec_notations}

\begin{center}
\begin{tabular}{ll}
\toprule
Abbreviations / Acronyms & Definition \\
\midrule
%BIL100 & \textcolor{red}{XXXX} \\
BMU & Bayesian model updating \\
DT & digital twin \\
EDF & \'Electricité de France \\
%EP-RGL4 & \textcolor{red}{XXXX} \\
ESTICOL & Estimation du Colmatage \\
JNGV & Jumeau Numérique Générateur de Vapeur \\
NPCC & nuclear power plant component \\
NPWR & nuclear pressurized water reactor \\
PHM & Prognostics and Health Management \\
R \& D & Research \& Development \\
RUL & remaining useful life \\
SciML & Scientific Machine Learning \\
SG & steam generator \\
TSP & tube support plate \\
TVE & televised video examination \\
UQ & Uncertainty Quantification \\
VPCE & vector-values polynomial chaos expansion\\
\bottomrule
\end{tabular}
\end{center}

\section{Introduction}
\label{sec1}

\noindent The clogging of steam generators (SGs) in nuclear pressurized water reactors (NPWRs) is a degradation phenomenon affecting certain plants of the French nuclear fleet \citep{Prusek2013, Feng2023, Jaber2026-thesis}. This complex degradation process happens at the level of the tube support plates (TSPs) and arises from various factors, among which the most important are the flow accelerated corrosion of the secondary circuit pipes and heat exchangers and the accidental pollution from raw water entering the cooling circuit \citep{Prusek2013}. In the context of extending the operational lifetime of the French nuclear fleet, optimizing maintenance operations of the nuclear power plant components (NPPCs), such as the steam generator, are crucial. As a major actor of electricity production in Europe and the world, EDF has  identified this objective and is pursuing research through the “Steam Generator Digital Twin” ("Jumeau Numérique Générateur de Vapeur" - JNGV) R\&D project \citep{Deri2021}. The JNGV initiative aims to develop a digital platform that integrates physical modeling, statistical methods, and data assimilation to estimate and forecast clogging in SGs as well as other degradation phenomena such as fouling of SG tubes \citep{Prusek2012}. The long-term goal of this tool is to assist the EDF engineers in order to perform more informed maintenance, with the help of a visual platform collecting and summarizing all the information related to the current health state of the SG.\\

\begin{figure}[h!]
    \centering
    \includegraphics[width=1.0\linewidth]{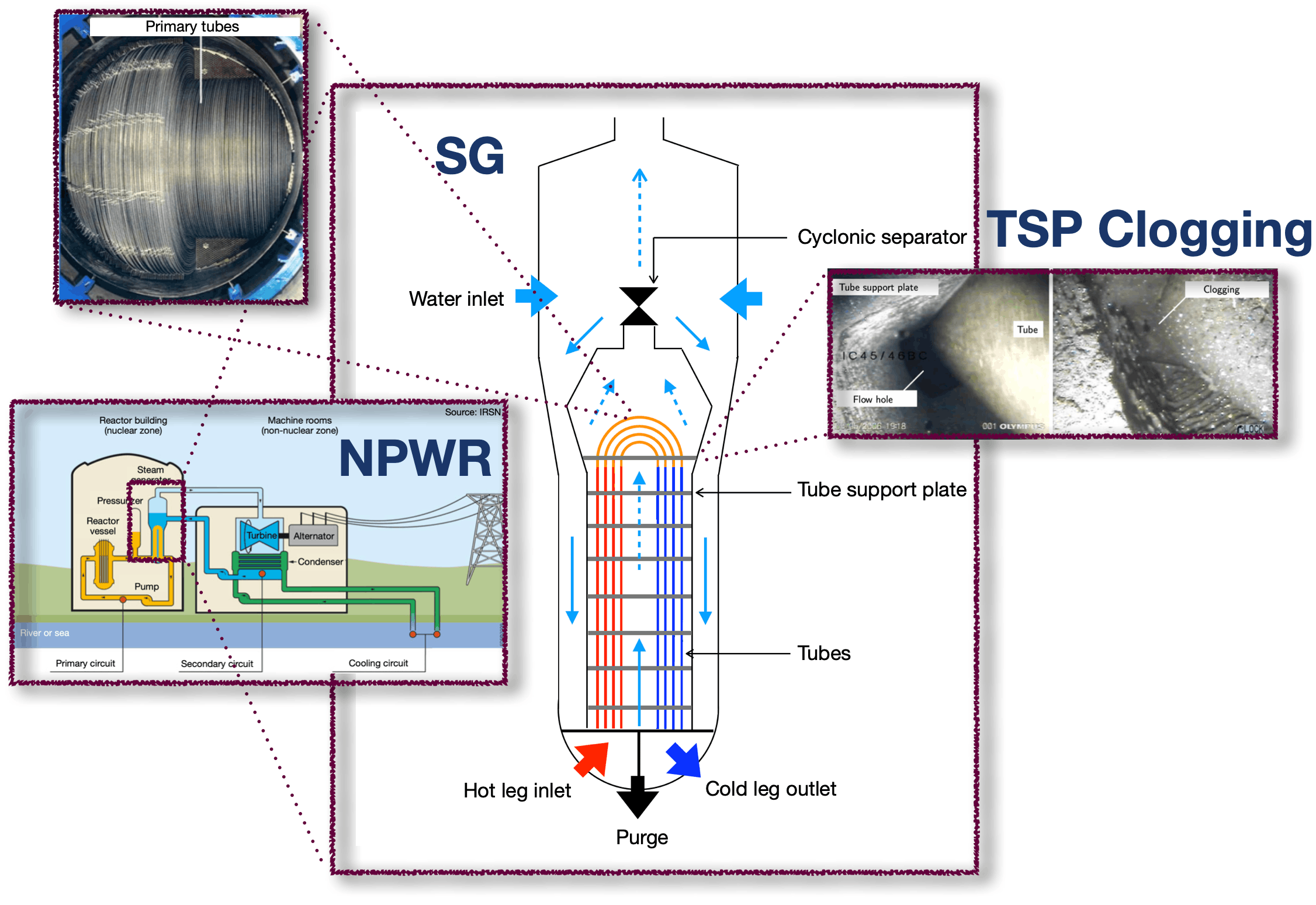}
    \caption{NPWR, SG and primary tube schemes, example of a TVE during a NPWR outage at the upper TSP (\textcopyright EDF, IRSN).}
    \label{fig:sg_tsp_clogging}
\end{figure}

\noindent Clogging is characterized by a clogging rate index denoted $\tau_{c}$ evaluated in the flow holes of the TSPs. It varies between $0\%$ and $100\%$ and is usually available for measurement only at the upper support plate of the SG, with the help of televised video examinations (TVEs) as shown in Figure~\ref{fig:sg_tsp_clogging}. It is known that the highest clogging rates \citep{Prusek2012} occur at the upper TSP. This degradation process is difficult and expensive to monitor because there is little feedback from the operating nuclear system until the clogging levels reach high thresholds, and because its own dynamics are slower compared to the whole lifetime of a SG. Moreover, data acquisition only occurs during plant shutdowns. For the latter, research is being undertaken concerning indirect acquisition methods using non-intrusive measurement techniques such as Foucault eddy currents or by finding other transient indicators \citep{McNab1988, Girard2014}. In addition, clogging has been observed to have a slow initiation kinetics, and due to the specific operational conditions of each NPWR, experimental testing in laboratory proves highly challenging. It is hard to build a clogging physical model due to the intertwined multiphysics interaction, notably involving fluid flows, heat transfer, chemistry of the secondary circuit, thermal-hydraulic evaporation conditions, complex component geometry and also the variety of operating conditions. Although it does not pose a major safety threat, clogging has diverse impacts on the SG, in particular, it can cause a localized redistribution of flow between the TSP flow holes (see Figure~\ref{fig:sg_tsp_clogging} below), which elevates the risk of flow induced vibration, the possible rupture of SG tubes, and hampers the SG response to operational changes. To tackle these challenges, preventive chemical cleaning maintenances can be performed, and these maintenances have to be planned optimally by estimating a reliable prognostics of the future clogging rate over the horizon of several months and even years. Chemical cleanings consist in stopping the NPWR and adding chemical solvents inside the SG that help disaggregate and remove part of the clogging deposits. \\

\noindent This maintenance planning problem lies at the core of the Prognostics and Health Management (PHM) framework \citep{Vachtsevanos2006}, whose purpose is to guide preventive maintenance actions in order to ensure reliable system performance. For a prescribed clogging degradation threshold $\tau_{c}^{*}$, prognostics consists in evaluating the \emph{Remaining useful life} (RUL) of the system \citep{Vachtsevanos2006}, defined as:
\begin{equation}
    \text{RUL}(t_{\chi c}, \tau_{c}^{*}) = \argmin_{\, t>t_{\chi c}} \{\tau_{c}(t) \ge \tau_{c}^{*} \},
\end{equation}
where $t_{\chi c}$ denotes the time of the last chemical cleaning.  
In practice, the evolution of $\tau_{c}(t)$ is influenced by a combination of deterministic phenomena such as physical laws (e.g. conservation laws and transport equations), and stochastic phenomena stemming from inherent variability in material properties, loading, and environmental conditions. Consequently, in all its generality, the degradation phenomenon must be treated probabilistically, and the RUL becomes a random variable corresponding to the first passage time at which the stochastic process $\tau_{c}(t)$ reaches threshold $\tau_{c}^{*}$ \citep{Escobet2019} (see Figure~\ref{fig:rul_proba} below). Prognostics therefore consists in estimating the probability distribution of this first passage time by means of a suitable degradation model in order to have a risk-informed decision strategy for maintenance planning. In this work, two classes of models are considered for clogging prognostics: \emph{physics-based} models based on physical models and implemented through numerical simulation, and statistical or \emph{data-driven models}, such as supervised learning-based approaches. 

\begin{figure}[h!]
    \centering
    \includegraphics[width=0.9\linewidth]{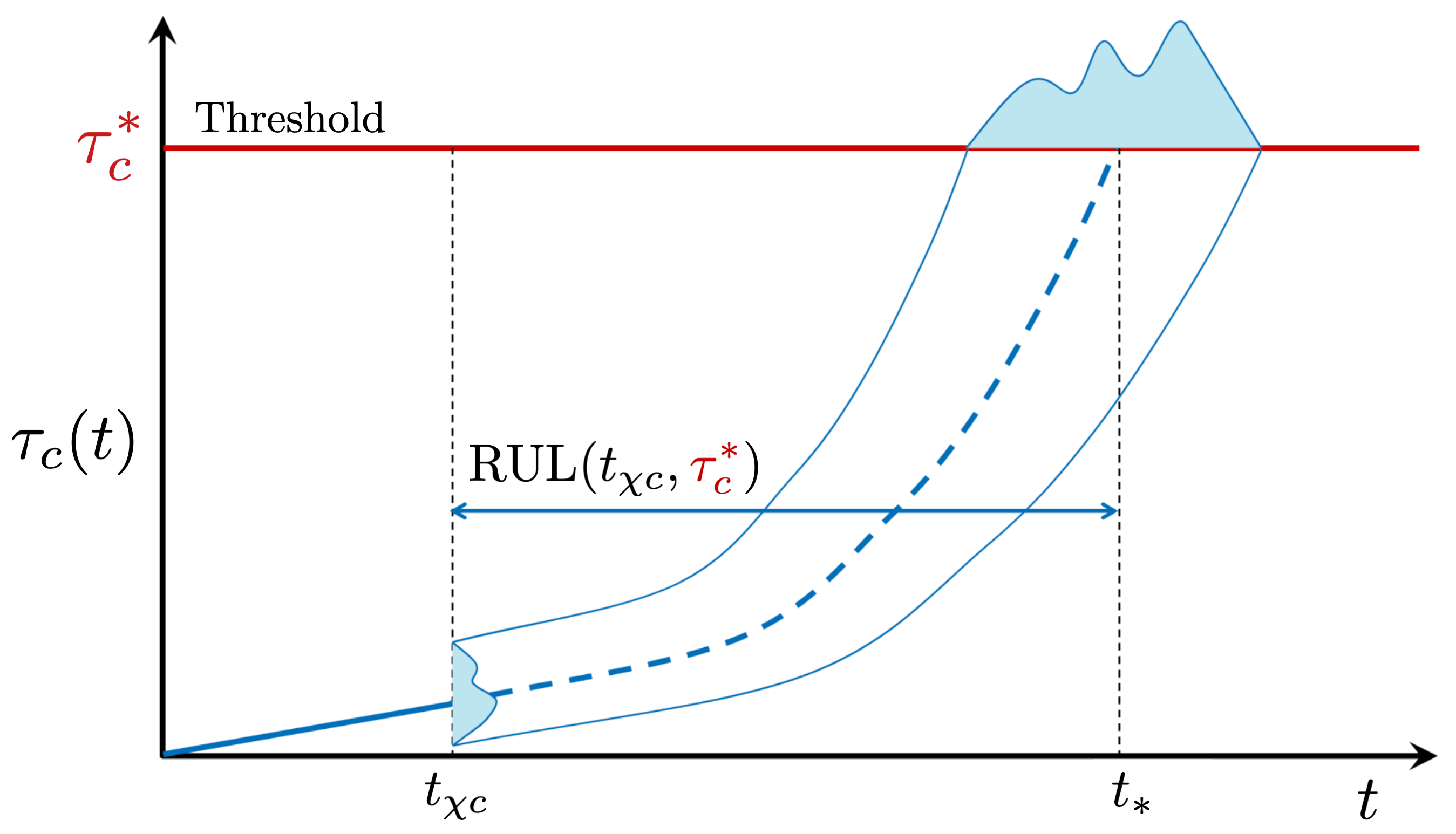}
    \caption{Prognostics process under uncertainty, adapted from: \citep{Escobet2019}}
    \label{fig:rul_proba}
\end{figure}

\noindent The central purpose of our work is to build so-called \emph{hybrid models} from these two approaches, a task which is at the core of so called Scientific Machine Learning (SciML) practices \citep{Rackauckas2020}, in order to guide maintenance planning using all the available clogging prognostics tools. The code idea of SciML is to leverage the complementary strengths of both physics-based and data-driven models to improve predictive accuracy, enhance generalization under limited data, and incorporate domain knowledge directly into learning pipelines, thereby yielding more stable and trustworthy estimates of the quantities of interest.\\

\noindent More specifically, the task here is to obtain robust probabilistic clogging RUL predictions with respect to uncertainties, by fusing all the available statistical information and numerical tools into a hybrid framework. This industrial challenge poses limitations to classical prognostics approaches, most notably because of the lack of substantial and consistent field data as well as the complex physics and the variety of operational conditions. Therefore one cannot hope to build purely data-driven generalizable algorithms and a purely physical, numerical approach that would be entirely trustworthy. Uncertainties are present in all the models and within this set of hypotheses, our goal will be to minimize the global prediction uncertainty and to guarantee as well as possible its control at every step in all its forms.\\

\noindent A more long-term objective of this work is to propose methodologies and frameworks for enabling digital twin (DTs) technologies of NPPCs \citep{NRC2021}. Digital twins are virtual representations of physical assets that integrate data, models, and simulations with associated uncertainties to enable real-time monitoring, diagnostics, and prognostics \citep{NAS2024, Liang2024}. In the context of NPWRs, DTs could support predictive maintenance, optimize operational strategies, and enhance safety by providing a continuously updated image of components' state of health. Achieving this requires the integration of physical models, data-driven approaches, uncertainty quantification (UQ) \citep{DeRocquigny2008}, and data assimilation techniques \citep{Evensen2022}. However, to date, there is still much to be done for obtaining real-time monitoring of relevant data on existing NPPCs, starting with sensor qualification and optimal placement \citep{Argaud2018} together with validation and verification of physical models and surrogates \citep{NRC2021, NRC2023}. The development of robust DTs relies on hybrid modeling frameworks that can adapt to new information, incorporate domain knowledge, and provide reliable predictions under uncertainty for existing facilities. As for the future and possibly renewed assets, there is room for better control at the design stage to facilitate the deployment of digital twin methodologies during the operation phase.\\

\begin{figure}[ht]
    \centering
    \includegraphics[width=1.0\linewidth]{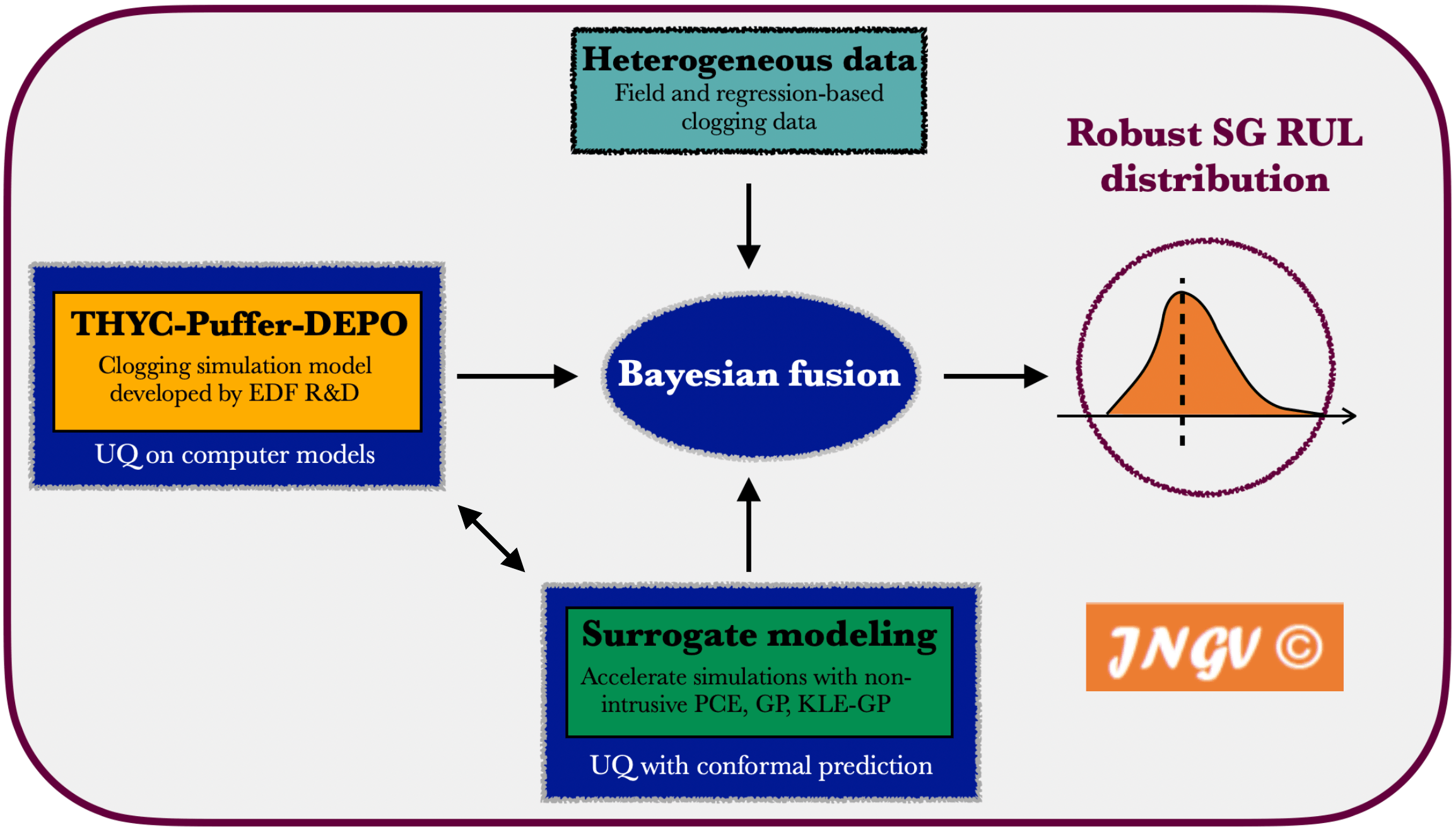}
    \caption{Hybrid framework for robust SG clogging prognostics.}
    \label{fig:SG_digital_twin}
\end{figure}

\noindent The remainder of this paper is organized as follows. In Section~\ref{sec1}, we provide an overview of DTs and discuss their application to NPPCs, with a particular focus on the SG clogging use case. Section~\ref{sec3} is devoted to a detailed presentation of the mathematical components of the hybrid framework shown in Figure~\ref{fig:SG_digital_twin}. We begin with the computer simulation code THYC-Puffer-DEPO (TPD) presented in Section~\ref{sec31}, then we describe the associated non-intrusive surrogate modeling techniques for accelerating the evaluation in Section~\ref{sec32}. Parallel to that, we also briefly describe different surrogate validation techniques including how the conformal prediction paradigm can be employed for UQ and subsequent model qualification. In Section~\ref{sec33}, we introduce the ESTICOL regression algorithm, which is used to enrich the existing database of SG clogging field measurements. Finally, Section~\ref{sec34} presents the core Bayesian methodology for fusing heterogeneous data sources, enabling robust clogging prognostics through input parametric uncertainty reduction in THYC-Puffer-DEPO code. Section~\ref{sec4} provides a conclusion and a discussion about some research perspectives for hybrid prognostics and DTs in the nuclear field.

\section{Presentation of the methodology}
\label{sec2}
\subsection{Digital twins for nuclear power plant components}
DTs are an emerging field of technologies \citep{NAS2024} that aims to bridge the gap between classical physical modeling and the growing amount of data in order to offer real-time guidance for informed decision-making on systems. The standard definition used in the different communities was given by the American Institute of Aeronautics and Astronautics \citep{AIAA2021} and is reproduced below:\\

\emph{\noindent A digital twin is a set of virtual information constructs that mimics the structure, context, and behavior of a natural, engineered, or social system (or system-of-systems), is dynamically updated with data from its physical twin, has a predictive capability, and informs decisions that realize value. The bidirectional interaction between the virtual and the physical is central to the digital twin.}\\

\noindent For NPWRs, this translates to designing systems maintaining continuous congruency with the actual state of the plant by adjusting in real-time, whenever necessary, to operational data. While still being at an early conceptual and design stage, guidelines and challenges for building such systems have been outlined in the following U.S Nuclear Regulation Committee report on DTs \citep{NRC2021}. The main features that NPPC-DTs must have is twofold. Firstly it must include modeling and simulation results, including data analytics, machine learning methods, physics-based models, and data-informed models (the latter meaning hybridizing physics-based models with real-time processing of plant data). Secondly it must have a robust data and information management system, covering accessibility, quality information and data storage as well as an operator interface with user-friendly visualization software. The report states that one of the primary purposes of NPPC-DTs is to support diagnostics and prognostics of the health state of structures, systems, enabling plant operators to anticipate failures and optimize preventive maintenance. The integration and use of such DT systems can save operating time and costs instead of relying only on failure observation and subsequent action since it mitigates potential unwanted maintenance costs coupled with shutdowns of the NPWR.\\

\begin{figure}[ht]
    \centering
    \includegraphics[width=1.0\textwidth]{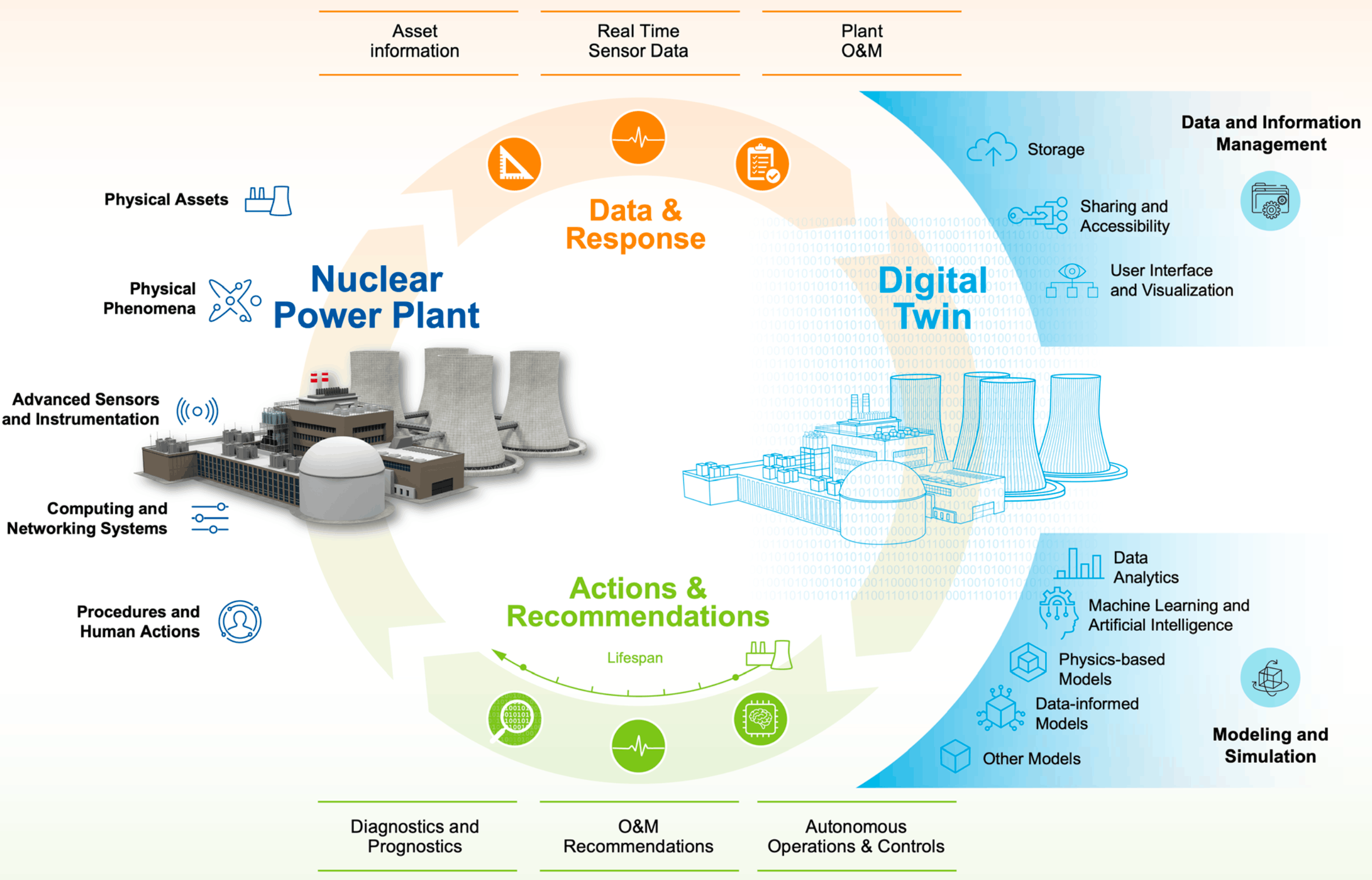}
    \caption{Summary of the workflow in a nuclear digital twin, \citep{NRC2021}.}
    \label{fig:nuclear_dt}
\end{figure}

\noindent While significant progress has been made, the adoption of DTs in the nuclear industry remains limited compared to sectors like civil engineering or renewable energies, primarily due to the specific challenges of nuclear prognostics. The field is still developing, but growing interest in the digitalization of future plant technologies is expected to accelerate research in this area. 

\subsection{Hybrid framework for the steam generator for clogging prognostics}
The purpose of the SG-DT (JNGV, see Section \ref{sec1}) is to provide an estimation of the health state of a specific SG and to better assist decision-making. It is supposed to give a virtual representation of the system by integrating all the available information on the specific system up to the present time. For the purpose of clogging prognostics, given that the clogging kinetics are slow and do not pose any real-time threat on the nuclear system, the DT must include a framework that does not require to be used online, but that has to be flexible enough to integrate any new arrival of data and to subsequently correct the predicted RUL distribution.\\

\noindent In order to get a robust RUL prediction of the SG-DT clogging prognostics, we propose a hybrid framework in which the computer simulation code TPD (detailed in Section~\ref{sec31}) acts as a base for the clogging prediction, and then quantify its parametric uncertainty coming from the poorly known of its input parameters. This prior uncertainty once evaluated, enables a first estimation of the RUL probabilistic distribution. During this UQ step, we build and use various non-intrusive surrogate modeling techniques in order to speed up the time-costly simulation code. However, for a proper use of these surrogates, they require careful and specific validation techniques such as what can be done with \emph{conformal prediction} \citep{Jaber2025-2}. Afterwards, we use the heterogeneous data coming from TVEs and the ESTICOL model (detailed in Section~\ref{sec33}) in a \emph{Bayesian fusion strategy}, allowing to reduce the state variable uncertainty \citep{Jaber2026} that we will present in Section~\ref{sec34}.  
\noindent For our specific use-case, the main challenges of the hybrid clogging prognostics can be summarized as follows:
\begin{itemize}
    \item \textit{Nonlinear, multi-physics modeling:} SG clogging involves complex, coupled transport, chemical deposition, and flow dynamics over multi-year timescales, requiring computationally intensive simulations with limited benchmarking \citep{Jaber2026-thesis}.
    \item \textit{Uncertainty quantification:} The parametric uncertainty of the computer simulation code must be rigorously assessed, serving as a baseline for uncertainty analysis in the hybrid prognostics framework, potentially leveraging surrogate modeling and conformal prediction.
    \item \textit{Sparse and heterogeneous measurements:} Observations of clogging rates are not frequent over the operational period and originate from diverse sources, including field inspections and regression-based estimates, complicating data integration.
    \item \textit{Offline prognostics assessment:} Data acquisition is not continuous or on-demand, necessitating offline assimilation strategies distinct from online filtering or sequential updating methods common in other PHM domains.
\end{itemize}

\noindent In summary, the hybrid prognostics problem for the SG clogging in NPPCs framework we have outlined, requires integrating physics-based simulation (THYC-Puffer-DEPO) with statistical models (such as ESTICOL) and field data (such as the TVEs). The main purpose is to produce robust, risk-informed RUL estimates despite sparse measurements and complex system dynamics, leveraging the complementary strengths of physical modeling and data-driven approaches within a digital twin framework.

\section{Presentation of the tools}
\label{sec3}
\subsection{Computer simulation model for clogging: THYC-Puffer-DEPO}
\label{sec31}
The clogging degradation studied here corresponds to the accumulation of iron oxide deposits on the boundaries of TSPs in SGs. The full long-term physical model of SG developed by EDF R\&D \citep{Prusek2013} clogging consists of three nested temporal levels of models, allowing simulation of the clogging rate $\tau_{c}$ over long periods of time (typically, from $40$ to $60$ years, and even more). First, the stationary thermal-hydraulic fields in the SG are computed. Using these fields, a transport mechanism carries the magnetite particles in both solid and soluble forms, which are likely to cause clogging following the \emph{vena contracta} giving rise to mass deposition through turbulent recirculation zones, and through \emph{flashing} mechanisms \citep{Prusek2013} (an additional electrokinetic phenomenon is also present but it is not modeled for the moment). Once the mass fractions from both phenomena are determined, a growth equation for the deposited magnetite mass is solved, enabling us to track the variation in the clogging rate at the different TSPs of the SG. More details on this model can be found in \citet{Prusek2013, Jaber2025-1}.\\

\noindent The simulation code THYC-Puffer-DEPO is based on the physical modeling mentioned above. It is implemented using a finite-volume, multiphysics approach allowing to simulate clogging degradation over the entire lifespan of the system. In brief, it is an association of three codes \citep{Feng2023}, \emph{THYC}: computing the stationary thermal-hydraulic two-phase fields inside the SG, \emph{Puffer}: computing the iron oxide solubility in the liquid phase, and finally \emph{DEPO}: which integrates the clogging model \citep{Prusek2013}, involving the vena contracta and associated flashing mechanisms. This computer code can be used for extrapolating the clogging dynamic on a specific SG to future times and therefore estimate the RUL for the system for a prescribed threshold. \\

\noindent Field experts have identified certain input variables that are subject to epistemic uncertainty and subsequent sensitivity analyses have been performed in \citet{Jaber2025-1} in order to assess the relative influence of these variables on $\tau_c$. Monte Carlo simulations have been enabled in order to evaluate the prior uncertainty by propagating uniform probability distributions over expert informed supports. These results can be seen in Figure~\ref{fig:sg_clogging_trajectories} below. Some additional informations can be seen on these plots, namely the occurrences of chemical cleanings (preventive or curative) as well as different chemical conditionings of the secondary circuit fluid using solvents $\chi_{i}$, $i\in\{1,2\}$ at different levels of pH (see \citet{Jaber2025-1} for more details), as well as the present time $t_{P}$ (fisplayed for illustrative purposes).  A first evaluation of the RUL density can be seen, and it is highly non-informative, given the last chemical cleaning time $t_{\chi c}$ because the trajectories have a high dispersion.

\begin{figure}[ht]
    \centering
    \includegraphics[width=1.0\linewidth]{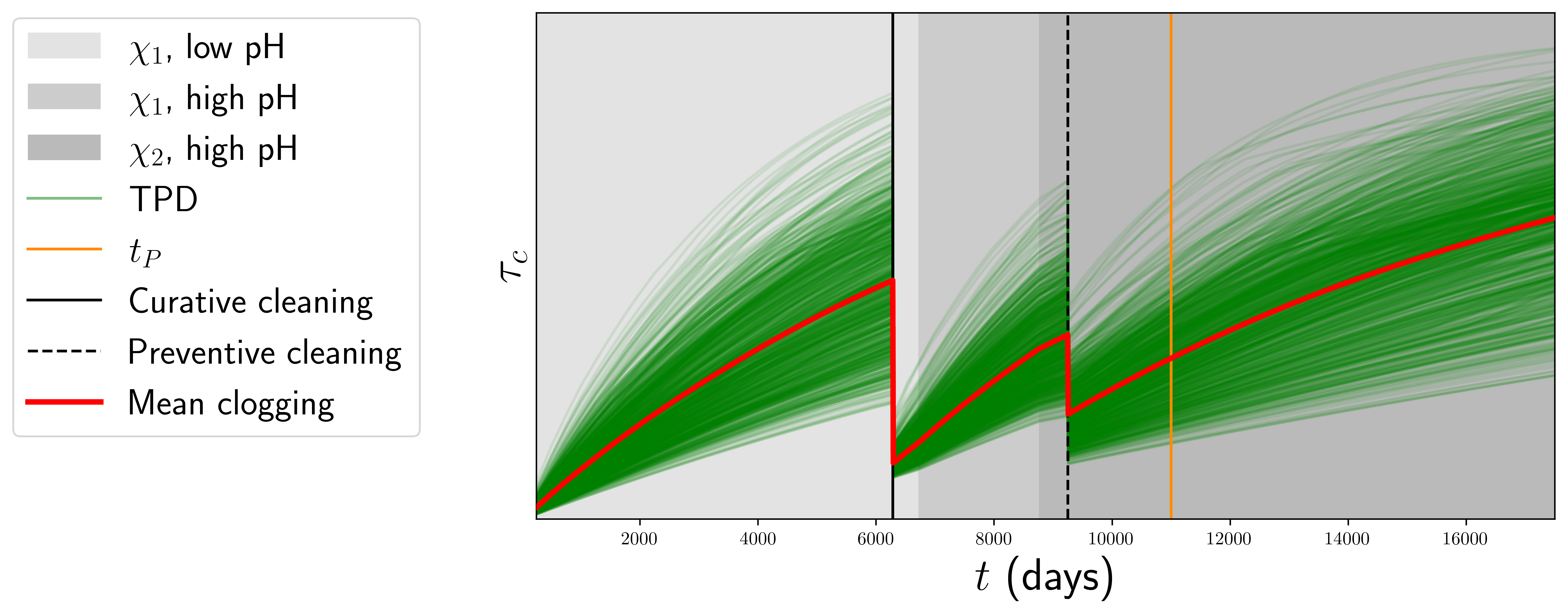}
    \caption{Clogging trajectories from uniform Monte Carlo simulations of TPD.}
    \label{fig:sg_clogging_trajectories}
\end{figure}

\subsection{Non-intrusive surrogate models of computer simulation codes}
\label{sec32}
A single run of the clogging simulation code requires approximately $5$~h on EDF high-performance computing infrastructure. Consequently, for applications involving a large number of model evaluations, such as Bayesian ensemble methods or other UQ and inference algorithms, the direct forward approach becomes computationally prohibitive \citep{DeRocquigny2008, Sullivan2015}. This motivates the construction of surrogate models based on methods borrowed from statistical learning. Using a first design of experiments (e.g. generated by Monte Carlo simulations or more advanced space-filling designs \citep{Ghanem2017}), standard statistical learning algorithms can be applied to approximate the model response and generalize to previously unseen inputs. The term \emph{non-intrusive} refers to the fact that the original simulation code is treated as a black box and is not modified, in contrast with intrusive approaches such as reduced basis methods based on Galerkin projections, where the numerical scheme itself is altered to accelerate computations \citep{Quarteroni2016}. Additional challenges arise from the nature of the model output, which in the present case is a time-dependent field rather than a scalar quantity. Common surrogate modeling approaches in this context include for instance Gaussian process regression \citep{Rasmussen2006}, polynomial chaos expansions \citep{Sudret2014}, support vector machines \citep{Bourinet2018}, artificial neural networks \citep{Hastie2009}. For THYC-Puffer-DEPO, multiple strategies can be used with equal good performances, in which case a more general approach involving surrogate aggregation can also be implemented \citep{Jaber2026}. In Figure~\ref{fig:PCE_TPD} below, an example is shown using a non-intrusive vector polynomial chaos expansions (VPCE) \citep{Jaber2025-1}. By computing the predictivity coefficient $Q^2$ \citep{Fekhari2023} at the different time instances of the simulation output, we see that it has a mean numerical value close to $1$, highlighting a good mean predictive performance. 

\begin{figure}[ht]
    \centering
    \includegraphics[width=1.0\linewidth]{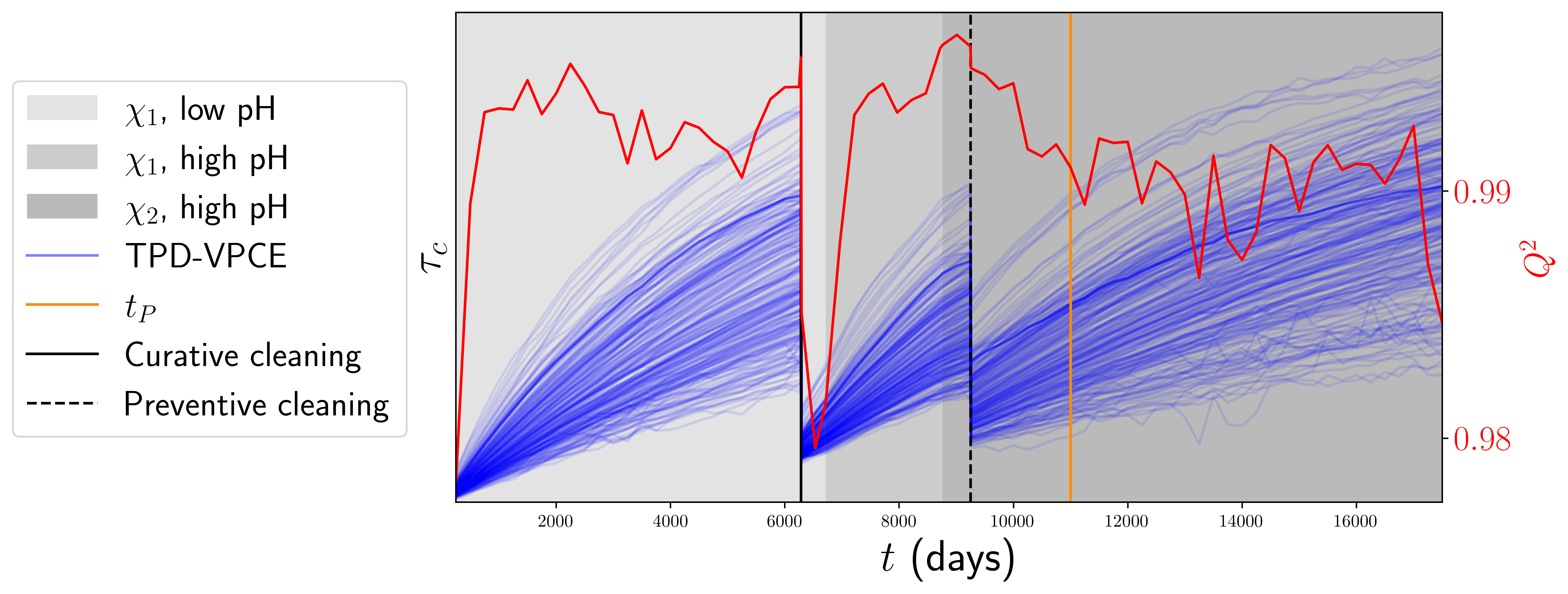}
    \caption{Non-intrusive vector polynomial chaos expansion \citep{Jaber2025-1} surrogate model of THYC-Puffer-DEPO and the time-variation of the predictivity coefficient $Q^2$.}
    \label{fig:PCE_TPD}
\end{figure}

\noindent As in any statistical learning strategy, surrogate modeling necessarily involves a validation step to assess the predictive performance of the approximation. Indeed, when such a model is used to evaluate quantities of interest derived from the reference code, for instance to estimate a RUL, the user must have at least a minimal knowledge of the associated risk. Metrics such as the predictivity coefficient \citep{Fekhari2023} are commonly evaluated and can be exploited for model selection. To achieve a robust UQ of these surrogate models, conformal prediction techniques may be employed \citep{Vovk2005}. These methods allow to estimate prediction intervals for the surrogate output at new input parameter values. In the case of non-intrusive surrogates for computer models, the underlying statistical learning framework makes it possible to apply methods such as split-conformal or cross-conformal approaches to construct prediction intervals. For Gaussian process surrogates, adaptive strategies can also be considered, such as the approach proposed in \citet{Jaber2025-2}. The resulting interval lengths provide an interpretable measure of the uncertainty and can further be used to discriminate between competing models according to the end-user’s risk tolerance. 

\subsection{Statistical model for clogging: ESTICOL}
\label{sec33}
The ESTICOL tool \citep{Pinciroli2021} is based on a regression approach that combines operational data with TVEs, which are visual inspections used to directly assess the clogging rate of SGs. A major difficulty in this application is that TVEs are rare: they can only be performed during NPWR outages, which occur at limited and irregular times over an asset’s operational life. As a result, for a given SG, only a very small number of TVE observations is typically available. \\

\noindent To mitigate the scarcity of direct observations, ESTICOL exploits information contained in standardized transient tests that are routinely performed for plant monitoring and safety purposes. These tests are conducted much more frequently than TVEs and provide operational signals that can be analyzed statistically. In particular, ESTICOL relies on the BIL100 in EP-RGL4 tests \citep{Pinciroli2021}, which are carried out according to well-defined procedures and documentation \citep{Paulin2008}. ESTICOL extracts descriptive features from these transient signals and correlates them with the available TVE observations in order to estimate past and current clogging levels \citep{Pinciroli2021}. An illustration of this method is provided in Figure~\ref{fig:esticol} below:\\

\begin{figure}[ht]
    \centering
    \includegraphics[width=1.0\linewidth]{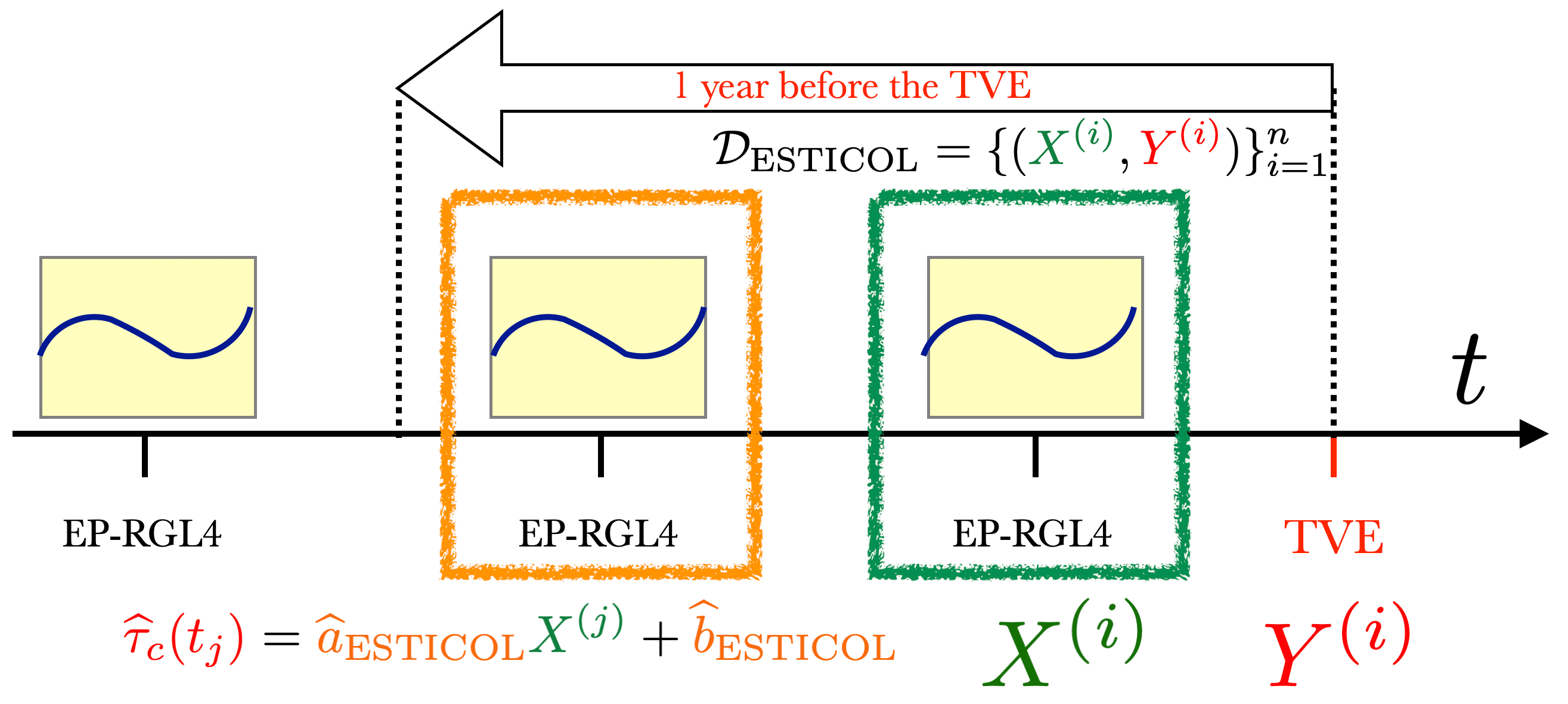}
    \caption{Idea of the ESTICOL algorithm \citep{Jaber2026-thesis}.}
    \label{fig:esticol}
\end{figure}

\noindent In practice, several limitations affect this approach. TVE observations remain extremely sparse and noisy, and the features derived from transient tests are subject to preprocessing and timing uncertainties. These factors can lead to variable predictive performance across SGs. Nevertheless, pragmatic validation strategies, such as cross-validation adapted to small data sets, indicate that ESTICOL can provide useful insights in favorable cases. Its main strength lies in offering low-cost, indirect estimates of clogging that can be combined with physics-based models to support robust assessment when direct inspection data are scarce.
The resulting enriched clogging data points together with the THYC-Puffer-DEPO trajectories are found in Figure~\ref{fig:trajectories_data_tpd_esticol_etv} below.

\begin{figure}[h!]
    \centering
    \includegraphics[width=1.0\textwidth]{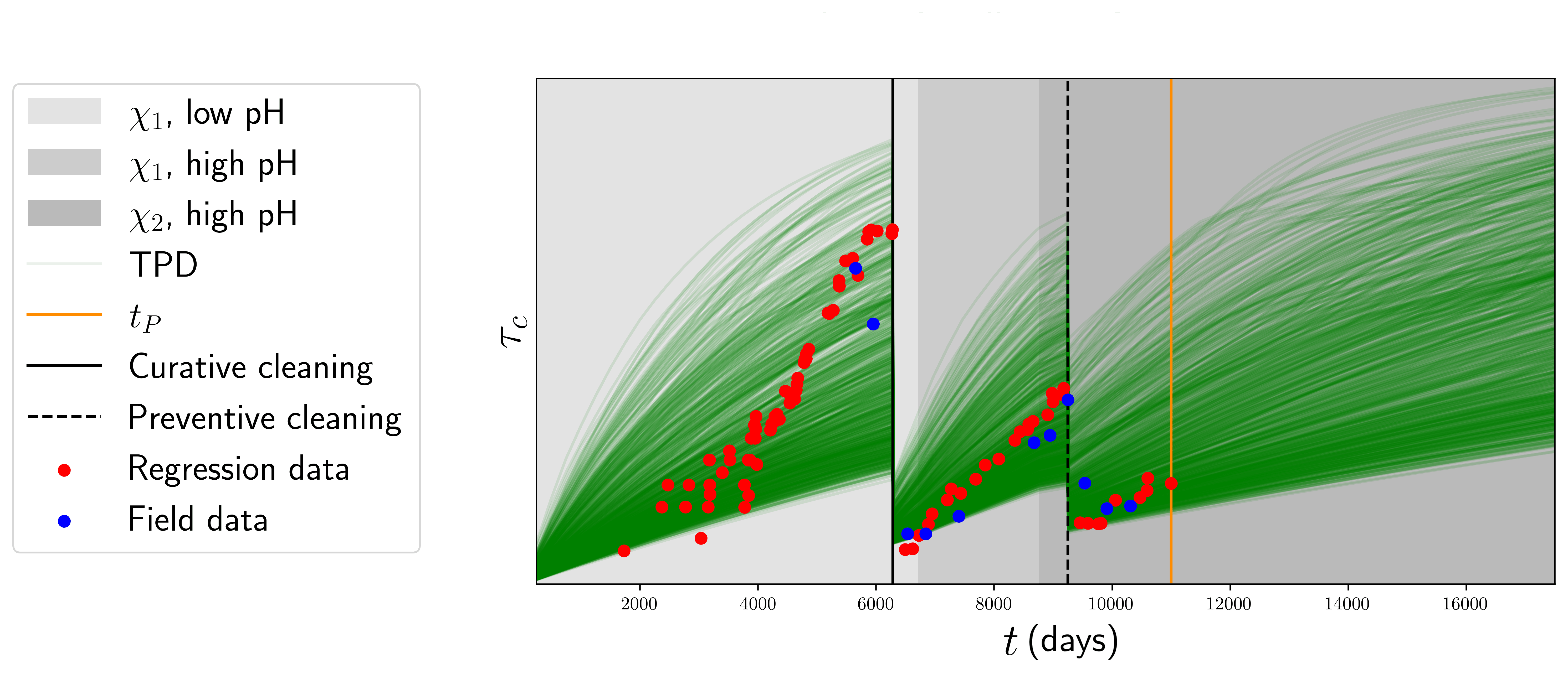}
    \caption{All the available data for the SG clogging prognostics problem. In red, the regression data points (ESTICOL) and field data points in blue (TVE).}
    \label{fig:trajectories_data_tpd_esticol_etv}
\end{figure}

\subsection{The proposed Bayesian fusion methodology}
\label{sec34}
We consider a data assimilation problem for degradation prognostics under uncertainty in a context where real-time degradation measurements are not available and only scarce and partial observations can be collected such as is the case for the clogging of SGs. We rely on a computer simulation model that takes a set of uncertain input parameters and produces, for each realization, a complete degradation trajectory over time. The model is treated as a grey-box: while the underlying physical mechanisms are known, the simulation code itself cannot be modified. Physical features of the solution, such as monotonicity are known. In addition to the simulation model, we consider several heterogeneous groups of degradation data originating from different sensors or statistical models and that have different levels of fidelity. Each data group is associated with its own observation times, and together they provide partial and irregular information on the same underlying degradation process. For each observation, we assume that the measured value corresponds to the true degradation level corrupted by a measurement noise whose variance is known a priori and that is specific to the data group. This is a working hypothesis and will be the topic of future developments.\\

\begin{figure}[h!]
    \centering
    \includegraphics[width=1.0\textwidth]{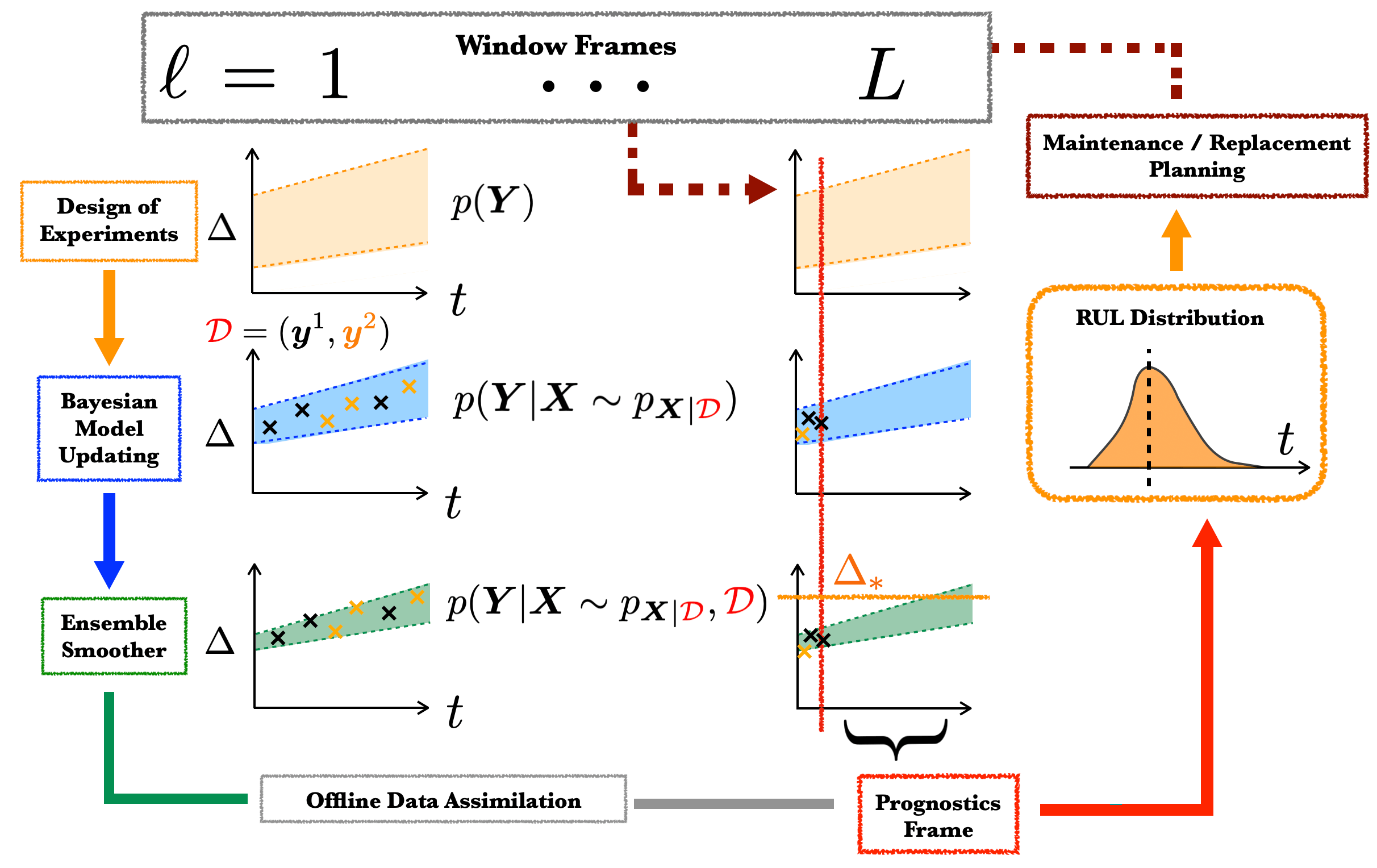}
    \caption{Proposed methodology for the offline data assimilation procedure \citep{Jaber2026}.}
    \label{fig:offline_hybrid}
\end{figure}

\noindent In the application considered here, the degradation index corresponds to the clogging rate $\tau_c$. The simulation model is the THYC-Puffer-DEPO code (see Section~\ref{sec31}), which depends on a vector of uncertain input parameters $\bm{X}$ and the associated surrogate models are used (see Section~\ref{sec32}) with output trajectories denoted by $\bm{Y}$. Two heterogeneous data sources are used: TVEs and regression-based estimates provided by ESTICOL (see Section~\ref{sec33}), gathered in a dataset $\mathcal{D} = \{\bm{y}^{\text{TVE}},\bm{y}^{\text{ESTICOL}}\}$ where each vector of points is of different size i.e $|\bm{y}^{\text{TVE}}|\neq |\bm{y}^{\text{ESTICOL}}|$.
The data assimilation procedure is performed sequentially over a series of predefined time windows. The overall methodology is summarized in Figure~\ref{fig:offline_hybrid}. For each time window, an initial set of simulations is generated using the original computer model. \emph{Bayesian model updating} (BMU) is then applied to revise the uncertainty on the input parameters based on the available information and obtain samples from the updated clogging trajectories distribution $p(\bm{Y}|\bm{X}\sim p_{\bm{X}|\mathcal{D}})$. The resulting ensemble of surrogate-approximated trajectories is subsequently conditioned on the heterogeneous data using an \emph{Ensemble Kalman smoothing} (EnKS) \citep{Evensen2000} technique to obtain samples from the full posterior $p(\bm{Y}|\bm{X}\sim p_{\bm{X}|\mathcal{D}},\mathcal{D})$. This step allows the available observations to be consistently incorporated into the model and provides a diagnostic check of the consistency between data and simulations. Once the prognostic horizon is reached, the same procedure is applied, and the distribution of the remaining useful life is finally estimated empirically. The mathematical details for this methodology are available in \citet{Jaber2026}. \\

\begin{figure}[ht!]
    \centering
    \includegraphics[width=1.0\textwidth]{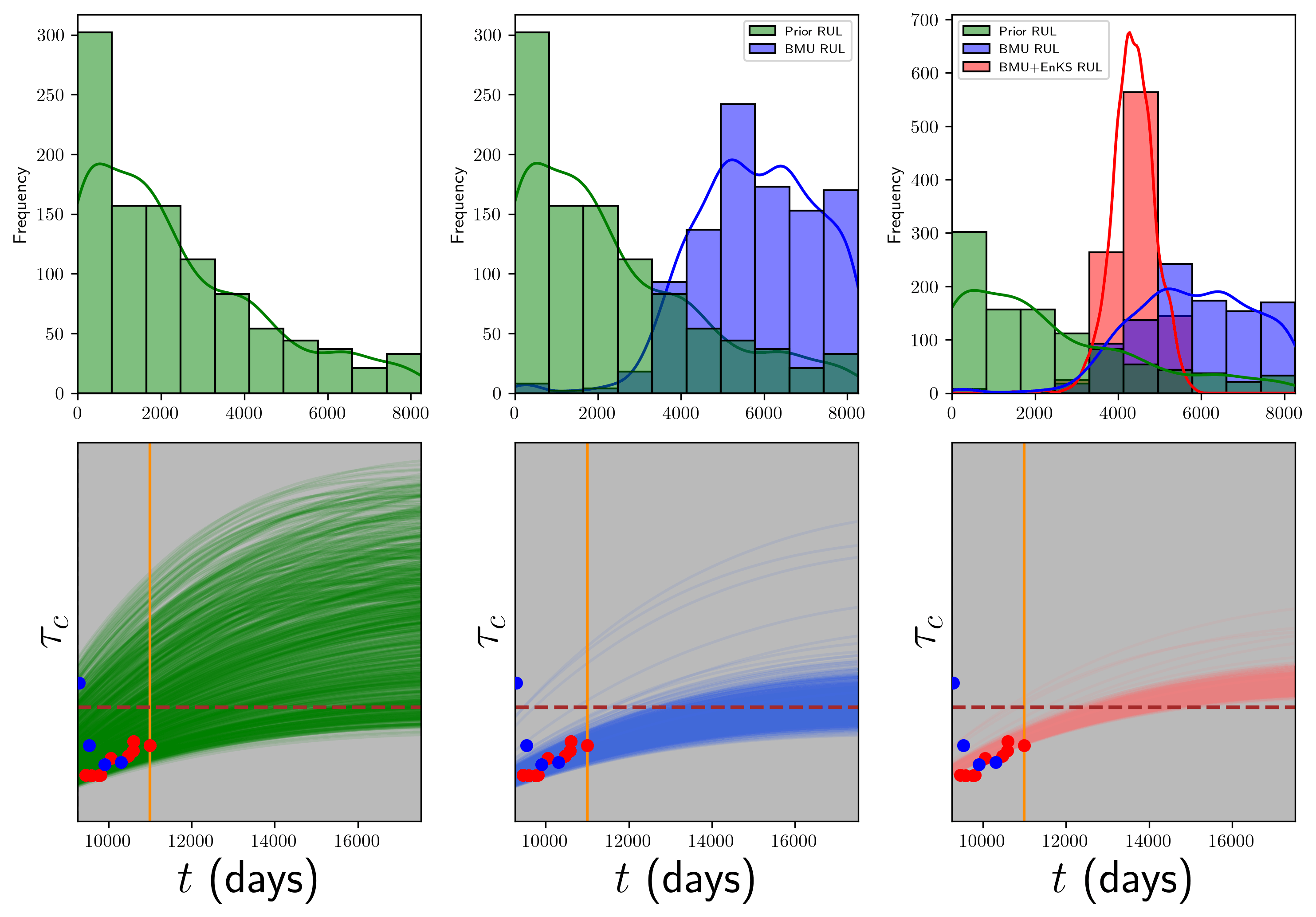}
    \caption{Prior and posterior RUL distributions before and after assimilation following the last preventive cleaning \citep{Jaber2026}, the yellow vertical line corresponds to the current operation time, the dashed horizontal line corresponds to the clogging threshold.}
    \label{fig:tpd_prior_posterior_ruls}
\end{figure}

\noindent Additionally, the methodology significantly improves the prediction of the RUL for a given threshold $\tau_{c}^{*}$. Figure~\ref{fig:tpd_prior_posterior_ruls} compares the prior and posterior RUL distributions after the last preventive maintenance (here taken as the prognostics time, the present time $t_{P}$ is displayed for illustrative purposes). The posterior RUL distribution is narrower and more concentrated at each step of the methodology, reflecting the reduced uncertainty and enhanced reliability of the predictions. This improvement is crucial for better maintenance planning and operational decision-making, as it provides a more accurate estimate of the time remaining before the next maintenance.

\section{Conclusion}
\label{sec4}
To summarize, we have developed a DT-based hybrid framework to better assist the prognostics of the clogging rate in flow holes of TSPs within NPWR SGs. This methodology employs a series of mathematical tools carefully selected for their methodological advantage (but knowing their own limitations), assembled within a hybrid method, with at its core a \emph{Bayesian fusion methodology of heterogeneous data}. A physical clogging simulation model (THYC-Puffer-DEPO) is used as a base prediction and its parametric uncertainty is reduced with the help of regression-enhanced database of operational data. The resulting trajectories can then be used to extrapolate at future times and allow to evaluate a robust and informed probabilistic RUL for different thresholds.\\

\noindent As a next step, the methods developed in this work could be integrated within EDF’s DT platform in order to facilitate their use by the EDF nuclear maintenance engineers. Extending the approach to the entire SG fleet affected by this degradation phenomenon would further demonstrate its operational relevance. Moreover, reliable DTs have the potential to deliver significant societal and industrial benefits. By improving predictive maintenance and reducing unplanned outages, they can enhance operational availability and support risk-informed decisions. In the nuclear context, this contributes not only to more efficient asset management but also to sustained public confidence and the ability to meet future energy needs.

\section*{Funding}
This work was sponsored by the ANRT-CIFRE Grant No. 2022/1412 as part of the PhD thesis of the first author.

\section*{Disclosure statement}
The authors declare no conflict of interest.
%%%%%%%%%%%%%%%%%%%%%%%%%%%%%%%%%%%%%%%%%%%%%%%%%%%%%%%%%%%%%%%%%%%%%%%%%%%%%%%

%%==================================%%
\bibliographystyle{sn-apacite}
\bibliography{References}
%%==================================%%

%%%%%%%%%%%%%%%%%%%%%%%%%%%%%%%%%%%%%%%%%%%%%%%%%%%%%%%%%%%%%%%%%%
%%%%%%%%%%%%%%%%%%%%%%%%%%%%%%%%%%%%%%%%%%%%%%%%%%%%%%%%%%%%%%%%%%
\end{document}